\documentclass[12pt]{article}

\usepackage{amsfonts}
\usepackage{amsmath}
\usepackage{amssymb}
\usepackage{here}
\usepackage{cite}                                                                                                                                                   
\usepackage{tikz}
\usetikzlibrary{matrix}
\usepackage{comment}                                                                                                                                                                            
\usepackage[labelformat=empty]{caption}

\newcommand{\ba}{\begin{eqnarray}}
\newcommand{\ea}{\end{eqnarray}}
\newcommand{\nn}{\nonumber}

\newcommand{\cF}{\mathcal{F}}
\newcommand{\cFs}{\hat{\mathcal{F}}}
\newcommand{\cH}{\mathcal{H}}

\newcommand{\cV}{\mathcal{V}}

\newcommand{\bR}{{\bf R}}
\newcommand{\bT}{{\bf T}}
\newcommand{\lra}{\leftrightarrow}
\newcommand{\rrangle}{\rangle}

\allowdisplaybreaks[1]

\begin{document}

\begin{titlepage}                                                                                                                                                                 
                                                                                                                                                                                               
\begin{flushright}                                                                                                                                                                             
UT-15-12                                                                                                                                                                                     
\end{flushright}                                                                                                                                                                               
                                                                                                                                                                                              
\vskip 12mm                                                                                                                                                                                    
                                                                                                                                                                                               
\begin{center}                                                                                                                                                                                 
{\Large Yangian associated with 2D ${\cal N}=1$ SCFT}                                                                                                                                                                                 
\vskip 2cm                                                                                                                                                                                     
{\Large Rui-Dong Zhu and Yutaka Matsuo}                                                                                                                                                                               
\vskip 2cm                                                                                                                                                                                     
{\it Department of Physics, The University of Tokyo}\\
{\it Hongo 7-3-1, Bunkyo-ku, Tokyo 113-0033, Japan}                                                                                                                                                                               
\end{center}                                                                                                                                                                                   
\vfill                                                                                                                                                                                         
\begin{abstract}                                                                                                                                                                               
Recently, Maulik and Okounkov proposed an integrable lattice model
where the degree of freedom at each site is identical to the Hilbert space
of free boson in two dimensions.  We give a brief review of their construction
and explain the relation with ${\cal W}_n$ algebra and Calogero-Sutherland
model.
As a generalization, we examine the Yangian associated with ${\cal N}=1$ superconformal
algebra which describes a supersymmetric extension of Calogero-Sutherland model
and compare it with the literature.
\end{abstract}                                                                                                                                                                                 
\vfill                                                                                                                                                                                         
\end{titlepage}                                                                                                                                                                                
                                                                                                                                                                                               
\setcounter{footnote}{0}


\section{Introduction}
A few years ago, Alday, Gaiotto and Tachikawa proposed a conjecture
\cite{Alday2010}
that the partition function of 4D ${\cal N}=2$ supersymmetric gauge theories
is identical to the correlation function of Liouville (Toda) system
in 2D.  It triggered rapid development of the technology of
2D CFT and in a few years, there appeared some proofs of the claim
in \cite{Alba2011, Fateev:2011hq, schiffmann2013cherednik} 
and a detailed study of the mathematical foundation was
carried out in \cite{Maulik:2012wi}.

Some features of  \cite{schiffmann2013cherednik, Maulik:2012wi} 
are the use of techniques of the integrable models
such as the Calogero-Sutherland system and
the symmetry behind it.  In \cite{schiffmann2013cherednik},
they used a nonlinear symmetry, degenerate double affine Hecke algebra
which is a natural symmetry of Calogero-Sutherland.
One may regard three objects, the cohomology ring of Hilbert scheme of
points on surfaces ($U(1)$ instanton), the Calogero-Sutherland model
and free boson in 2D as the same objects.  
All of them are characterized by a Hilbert space whose basis is labeled
by a Young diagram $Y$,  Jack polynomial in case of the Calogero-Sutherland,
the Fock basis for the free boson,
and  the fixed points of torus action for the equivariant
cohomology of Hilbert scheme.

For the description of $SU(N)$ gauge theories, one has to extend the framework
to include the basis with $N$ Young tables.  We have to compose the
states labeled by single $Y$ to give those with $N$ $Y$s.
For the free boson it is defined by the quantum Miura transformation
which defines ${\cal W}_n$-algebra \cite{Fateev:1987zh}.  For the Calogero system,
it is given by a coproduct in spherical degenerate DAHA 
(refered as SH${}^{c}$)\cite{schiffmann2013cherednik} which defines
the algebraic structure among the Jack polynomials.
A more explicit construction of the coproduct was presented in \cite{Maulik:2012wi}
where they introduced the structure of integrable lattice model
to analyze the problem. 

A new feature of the last approach is to use an
infinite dimensional Fock space $\cF$ generated by Virasoro operators 
on each of the lattice site
as the internal degree of freedom.\footnote{We note that
a similar integrable model was studied \cite{Bazhanov:1994ft, Bazhanov:1996dr, Bazhanov:1998dq}
during 90's where an infinite set of commuting charges were obtained in terms of the vertex operators.
The construction of R-matrix is different from \cite{Maulik:2012wi} 
and so is the integrable model thus obtained.
}
It seems to give a new inspiration in the definition of 2D CFT.
In this paper,  as the simplest application of their idea, we define
an integrable lattice model by
${\cal N}=1$ superconformal field theory.  It gives an infinite number
of commuting charges by the combination of bosonic and fermionic oscillators.
It naturally defines a supersymmetric generalization of Calogero-Sutherland model
which may be compared with the known results 
in the literature \cite{Desrosiers:2001ri}.
Furthermore, the higher generators may define a supersymmetric ${\cal W}$ algebra \cite{Inami:1988xy}.

We organize the paper as follows.  In section \ref{s:Yangian},
we briefly review the integrable lattice model proposed in \cite{Maulik:2012wi}.
We explain the definition of R-matrix and the construction of the commuting charges.
In particular, for the simplest case where the lattice consists of a point, we obtain
the Hamiltonian of the Calogero-Sutherland system as one of the conserved charges.
For the lattice with $N$ sites, the monodromy matrix gives the definition of associated Yangian.
It is defined in terms of $N$ bosons and is equivalent to a combined system with
${\cal W}_N$-algebra + $U(1)$ boson.
In section \ref{s:susy}, we consider a
generalization to $\mathcal{N}=1$ SUSY case.  We define the R-matrix in terms
of the commutation relation of superconformal generators
as a reflection operator.  The construction guarantees the existence
of infinite number of commuting charges.  It defines a supersymmetric
generalization of Calogero-Sutherland system when the number of lattice is one.
For the general cases, the Yangian is defined in terms of $N$ pairs of boson and
fermion fields.  We compare the second commuting charge with the Hamiltonian
of super Calogero-Sutherland model discussed in \cite{Desrosiers:2001ri}
and find some discrepancies. It may suggest that the supersymmetric extension of Calogero-Sutherland
system is not unique.

\section{Yangian for 2D CFT}\label{s:Yangian}
In this section, we review the construction 
of the integrable lattice model with a Fock space on each lattice site, 
proposed by Maulik and Okounkov (MO) \cite{Maulik:2012wi}. 
The Fock space is generated by a standard free boson in 2D: 
\ba
&&\phi(z)=\rho_0+\alpha_0 \log z-\sum_{n\neq 0} \frac{\alpha_n}{n} z^{-n}\,,\\
&& [\alpha_n, \alpha_m]=n\delta_{n+m,0}\,,\qquad [\alpha_0, \rho_0]=1\,.
\ea
The vacuum state $|\eta\rangle$ is defined by
$\alpha_n |\eta\rangle=0$ ($n>0$) and $\alpha_0|\eta\rangle=\eta|\eta\rangle$
and we denote  the Fock space generated from it as $\cF(\eta)$.
We are going to define a lattice model where 
we have an independent  Fock space $\cF(\eta_i)$ on each site $i$
($i=1,\cdots N$).
The boson field which acts on $\cF(\eta_i)$ will have extra index as
$\phi^{(i)}, \alpha^{(i)}_n$ when the distinction is necessary.

\subsection{Definition of R-matrix}
In the integrable models, the R-matrix is defined as operator acting on the
direct product of linear spaces, say $V$: $\bR(u): V\otimes V\longmapsto V\otimes V$.
Here we use the Fock space $\cF(\eta_i)$ as such vector space. 
We write two Fock spaces as $\cF(\eta_i)$ ($i=1,2$) in the following.
R-matrix is an endmorphism ${\bf R}(u):{\cal F}(\eta_1)\otimes{\cal F}(\eta_2)
\longmapsto{\cal F}(\eta_1)\otimes{\cal F}(\eta_2)$.
The rapidity parameter $u$ is related to 
the difference of $\eta$,  $u:=(\eta_2-\eta_1)/\sqrt{2}$.
We write $\phi^\pm=\frac{1}{\sqrt{2}}\left(\phi^{(1)}\pm\phi^{(2)}\right)$
and $\alpha_n^\pm=\frac{1}{\sqrt{2}}(\alpha^{(1)}_n\pm \alpha^{(2)}_n)$.

MO's R-matrix is defined to depend only on
$\phi^{-}$ but not on $\phi^+$.
We introduce the Virasoro operator for $\phi^-$ as,
\ba
L_n(u, \rho)=\frac{1}{2}\sum_m{}':\alpha^-_{n+m}\alpha^-_{-m}:+\rho n\alpha^-_n-
u \alpha^-_n=: L_n^{(0)}+\rho n\alpha^-_n-u\alpha_n^-\,.\label{MO}
\ea
$\sum{}'$ denotes the summation without terms involving zero mode and 
$L_n^{(0)}:=\frac{1}{2}\sum_m{}':\alpha^-_{n+m}\alpha^-_{-m}:$.
We note that we have replaced $\alpha^-_0$ by its eigenvalue
 on $|\eta_1\rangle_1\otimes |\eta_2\rangle_2$.
$L_n(u, \rho)$ satisfies Virasoro algebra with central charge $c=1-12\rho^2$ except for
a constant shift of $L_0$.\footnote{
We note that careful readers might notice that the coefficient of $\alpha^-_{n}$ is $\rho n$
instead of the standard one $\rho (n+1)$.  It comes from the fact that we use
the coefficients in $w$ patch instead of the $z$ patch ($z=\exp(w)$).
}

We introduce four operators $R_{\epsilon_1, \epsilon_2}$ ($\epsilon_i = \pm$)
which act on the Fock space $\cF^-(u)$, the Hilbert space for $\phi^-$:
\begin{eqnarray}
R_{\epsilon_1,\epsilon_2}(u,\rho)L_{n}(u, \rho)& =&
L_{n}(\epsilon_1 u, \epsilon_2 \rho) R_{\epsilon_1,\epsilon_2}(u,\rho)\label{ree1}\,,\\
R_{{\epsilon_1,\epsilon_2}}(u,\rho)|u\rangle^-&=&|\epsilon_1 u\rangle^-\,,\label{ree2}
\end{eqnarray}
where $|u\rangle^-$ is the Fock vacuum for $\phi^-$ with $\alpha^-_0|u\rangle^- = -u|u\rangle^-$.
Obviously the operator $R_{++}=1$ is trivial.
The operator $R_{--}$ is characterized by
relations $\alpha_n R_{--}=-R_{--}\alpha_n$ 
which can be easily solved.
$R_{-+}$ is a nontrivial operator which reverses the momentum 
direction and is known as the reflection operator in Liouville theory
\cite{Teschner:2001rv}. 
$R_{+-}$ is a novel operator which is the focus of this paper.
We will see later that the conditions (\ref{ree1},\ref{ree2}) uniquely determines $R_{\epsilon_1,\epsilon_2}$. 
We also note that in \cite{Shou:2014pxa} the reflection
operator plays essential role in the construction of AFLT
states \cite{Alba2011,Fateev:2011hq}.

The R-matrix is defined as $\bR(u):=R_{+-}(u,\rho)$.
The conditions (\ref{ree1}, \ref{ree2}) are replaced
by:
\ba
&& {\bf R}(u) L_{n}(u, \rho)=L_{n}(u, -\rho )\bR(u) \label{R-def}
\\
&& \bR(u)|\eta_1\rangle\otimes|\eta_2\rangle=|\eta_1\rangle\otimes|\eta_2\rangle\,.
\label{R-def2}
\ea

In \cite{Maulik:2012wi}, it was shown that this R-matrix satisfies the Yang-Baxter equation (YBE).
Instead of following their argument, we give a shortcut of the proof which depends only on
the definition in terms of Virasoro generators.
For the discussion of Yang-Baxter relation, we introduce
a new operator $\check{{\bf R}}(u):= {\cal P}{\bf R}(u)$, 
where ${\cal P}$ is the operator exchanging two Fock spaces $\cF_1, \cF_2$. 
On the $\phi^-$ subspace, $\check{{\bf R}}(u)=R_{-+}$, while on the $\phi^+$ subspace, ${\bf R}(u)$ acts trivially. 
The commuting chart (the YBE) for $\check{{\bf R}}(u)$ is\\
\begin{tikzpicture}
  \matrix (m) [matrix of math nodes, row sep=3em,column sep=4em,minimum width=2em]
  {
     {\cal F}(\eta_1)\otimes {\cal F}(\eta_2)\otimes{\cal F}(\eta_3) &  {\cal F}(\eta_2)\otimes {\cal F}(\eta_1)\otimes{\cal F}(\eta_3)
& {\cal F}(\eta_2)\otimes {\cal F}(\eta_3)\otimes{\cal F}(\eta_1)\\
     {\cal F}(\eta_1)\otimes {\cal F}(\eta_3)\otimes{\cal F}(\eta_2) & {\cal F}(\eta_3)\otimes {\cal F}(\eta_1)\otimes{\cal F}(\eta_2)
& {\cal F}(\eta_3)\otimes {\cal F}(\eta_2)\otimes{\cal F}(\eta_1)\\};
  \path[-stealth]
    (m-1-1) edge node [left] {$\check{\bf R}_{23}(u_3)$} (m-2-1)
            edge node [above] {$\check{\bf R}_{12}(u_1)$} (m-1-2)
    (m-2-1) edge node [above] {$\check{\bf R}_{12}(u_2)$} (m-2-2)
    (m-1-2) edge node [above] {$\check{\bf R}_{23}(u_2)$} (m-1-3)
    (m-2-2) edge node [above] {$\check{\bf R}_{23}(u_1)$} (m-2-3)
    (m-1-3) edge node [right] {$\check{\bf R}_{12}(u_3)$} (m-2-3);
\end{tikzpicture}\\
\noindent where $u_1=\frac{\eta_2-\eta_1}{\sqrt{2}}$, $u_2=\frac{\eta_3-\eta_1}{\sqrt{2}}$ 
and $u_3=\frac{\eta_3-\eta_2}{\sqrt{2}}$. The relation $u_3=u_2-u_1$ with the chart above shows the YBE.
We can see that the background charge of the $\phi^-$ subspace 
serves exactly as the spectral parameter (rapidity parameter).

For later reference, let us write down the YBE for ${\bf R}(u)$. 
\ba
{\bf R}_{12}(u-u'){\bf R}_{13}(u){\bf R}_{23}(u')={\bf R}_{23}(u'){\bf R}_{13}(u){\bf R}_{12}(u-u')\label{YBE}
\ea

\subsection{Explicit Computation of ${\bf R}(u)$}
\ \ \ We note that when $u\rightarrow\infty$, all terms involving 
$\rho$ can be neglected, and therefore ${\bf R}(u)$ behaves as the identity operator under this limit. 
The R-matrix, therefore, can be expanded as follows
\ba
{\bf R}(u)=R_{+-}(u,\rho)\equiv \sum_{n=0}^\infty R^{(n)} u^{-n}
\equiv\exp\left(\sum_{n>0}\frac{r^{(n)}}{u^n}\right)\,.\label{R-exp}
\ea
The relation between $R^{(n)}$ and $r^{(n)}$ is expressed for the lower orders as,
\ba
R^{(0)}&=&1\nonumber\\
R^{(1)}&=&r^{(1)}\nonumber\\
R^{(2)}&=&r^{(2)}+\frac{1}{2}r^{(1)}r^{(1)}\nonumber\\
&\dots &\nonumber
\ea
The defining recursion equation for ${\bf R}(u)$ can then be expressed as
\ba
[R^{(m)},\alpha^-_n]=[R^{(m-1)},L_n^{(0)}]+\rho n\{R^{(m-1)},\alpha^-_n\}\,.\label{recursion}
\ea
In general (\ref{R-exp}) may be solved from the lower orders.
It may not be so obvious if such a system of equations is
consistent.  The Jacobi identity serves as such consistency condition,
\ba
[[R^{(i)},\alpha^-_n],\alpha^-_m]=[[R^{(i)},\alpha^-_m],\alpha^-_n].\label{Jacobi}
\ea
The confirmation of such identity is straightforward but a bit lengthy.
We give some explicit computation in appendix \ref{s:Jacobi}.

Let us demonstrate the explicit form of R-matrix for the lower orders.
The recursion formula can be written more explicitly in terms of $r^{(i)}$  as,
\ba
\left[r^{(1)},\alpha^-_n\right]&=&2\rho n\alpha^-_n\,,\label{r-1}\\
\left[r^{(2)},\alpha^-_n\right]&=&\rho nr^{(1)}\alpha^-_n+\rho n\alpha^-_nr^{(1)}+[r^{(1)},L_n^{(0)}]-\frac{1}{2}[\left(r^{(1)}\right)^2,\alpha^-_n]\label{r-2}\,,\\
\dots\nonumber
\ea
These equations leave a degree of freedom of adding constants to $r^{(n)}$'s, 
which will be fixed by condition 
(\ref{R-def2}).
Eq. (\ref{r-1}) can be solved to
\ba
r^{(1)}=-2\rho L^{(0)}_0=-2\rho\sum_{n>0}\alpha^-_{-n}\alpha^-_n
\ea
Substituting (\ref{r-1}) into (\ref{r-2}), we get
\ba
[r^{(2)},\alpha^-_n]=[r^{(1)},L_n^{(0)}]=2\rho nL_n^{(0)}\,.\nonumber
\ea
It is solved as,
\ba
r^{(2)}=-\frac{2}{3}\rho\sum_{n}{}':L_n^{(0)}\alpha^-_{-n}:=-\frac{1}{3}\rho\sum_{n,m}:\alpha^-_{n+m}\alpha^-_{-m}\alpha^-_{-n}:\nonumber\\
=-\rho\sum_{n,m>0}\left(\alpha^-_{-n}\alpha^-_{-m}\alpha^-_{n+m}+\alpha^-_{-n-m}\alpha^-_n\alpha^-_m\right)
\ea
The higher coefficients  $r^{(3)}, r^{(4)}$ will be given in Appendix \ref{s:r-3}.

\subsection{Yangian}
As a standard strategy in the integrable models, one can construct
the monodromy operator ($T$-matrix) through the R-matrix.
We use the R-matrix as the $L$-operator and treat the Fock space $\cF$
(say $\cF_0$) as the auxiliary space.  The $T$-matrix is defined as,
\ba\label{Tmat}
\bT_0(u):= {\bf R}_{01}(u){\bf R}_{02}(u)\dots{\bf R}_{0N}(u)\,,
\ea
which defines an endmorphism on $\cF_0\otimes(\otimes_{i=1}^N \cF_i)$.
The $T$-matrix satisfies the fundamental algebraic relation,
\ba
\bR_{00'}(u-u')\bT_0(u) \bT_{0'}(u')=\bT_{0'}(u') \bT_0(u) \bR_{00'}(u-u')\,.
\ea

We write  a brief review of Yangian symmetry in the appendix \ref{a:Yangian}.
The $T$-matrix defined here corresponds to $T_{ab}(u)$
in (\ref{Tab}). To make the conncection clearer, let us rewrite the monodromy matrix as
\ba
T_{ij}(u)=\delta_{ij}+\frac{h}{u}\sum_{J}\sum_a t^a_{ij}I^a_J+\frac{h^2}{u^2}\sum_{J<K}\sum_{a,b}\sum_{k}t^a_{ik}t^b_{kj}I^a_JI^b_K+\dots\,,
\ea
where $a$ is the adjoint index of $su(p)$ and $I^a$ is a generator of this Lie algebra. The combination $t^a_{ij}I^a_J$ acts on the tensored space 
$\mathbb{C}^p_0\otimes V_J$ with $t^a_{ij}$ acting on the first vector space and $V_J$ is the representation space for $I_J$'s. 
The associated R-matrix can be written as (up to a rescaling)
\ba
R(u)=1\otimes1-\frac{h}{u}\sum_{a,b}g_{ab}t^a\otimes t^b\,,
\ea
with the metric $g_{ab}\propto{\rm tr}(t^at^b)$ is an invariant inner product of the algebra. 
For more general Lie algebras, Belavin and Drinfeld \cite{Belavin-Drinfeld:1982} showed that the asymptotic form of the R-matrix is 
\ba
R(u)=1\otimes1-\frac{h}{u}\sum_{a,b}g_{ab}t^a\otimes t^b+{\cal O}(\frac{h^2}{u^2})\,,
\ea
and the Yangian is uniquely determined. 
We can immediately see the correspondence: $T_{ab}$ is a product of operators $1+\frac{h}{u}S^{ab}_i$ on each site,
the corresponding operator is 
\ba\label{r0}
{\bf R}_{0i}(u) &=& 1 -\frac{\rho}{u}\sum_{n>0}( \alpha^{0}_{-n} -\alpha^{i}_{-n} )(\alpha_n^0-\alpha_n^i)+\dots\nn\\
&=&1-\frac{\rho}{u}\left(L_0\otimes 1+1\otimes L_0\right)+\frac{\rho}{u}\sum_{n>0}(\alpha_{-n}\otimes\alpha_n+\alpha_n\otimes\alpha_{-n})+\cdots\,.
\ea
which acts on ${\cal F}_0\otimes{\cal F}_i$ with $L_0=\sum_{n>0}\alpha_{-n}\alpha_n$. 
The underlying algebra spanned by $L_0$, $\alpha_n$'s and $1$ is the algebra for free boson.  
In \cite{Maulik:2012wi} it is refered as $\widehat{\mathfrak{gl}(1)}$
and the operators generated by the product (\ref{Tmat})
was defined as the Yangian of  $\widehat{\mathfrak{gl}(1)}$.
We note that in (\ref{r0}), the parameter $\rho$ plays the role of Plank constant $h$.
It is the deformation parameter that appears in the
Virasoro generators. 

It can be easily checked that the corresponding part of classical R-matrix
\ba
\mathfrak{r}=\frac{1}{u}\left(\sum_{n>0}(\alpha_{-n}\otimes\alpha_n+\alpha_n\otimes\alpha_{-n})-L_0\otimes1-1\otimes L_0\right)\label{classical-r}
\ea
satisfies the classical Yang-Baxter equation, 
\ba
[\mathfrak{r}_{12},\mathfrak{r}_{13}]+[\mathfrak{r}_{12},\mathfrak{r}_{23}]+[\mathfrak{r}_{13},\mathfrak{r}_{23}]=0\,,
\ea
where $\mathfrak{r}_{ij}=\mathfrak{r}_{ij}(u_i-u_j)$. This provides a consistancy check for the Yangian. The explicit computation is shown in Appendix \ref{s:cYBE}. 

The monodromy matrix can be formally expanded as,
\ba
\bT_0(u)=\sum_{\lambda,\nu} \vec\alpha^{(0)}_{-\lambda} \mathcal{O}_{\lambda\nu} (u)\vec{\alpha}^{(0)}_{\nu}
\ea
where the summation is over the Hilbert space $\cF_0$. 
Since it is the Fock space of free boson, one may parametrize it by
a Young diagram $\lambda$.  The operators $\vec\alpha^{(0)}_{-\lambda}, \vec{\alpha}^{(0)}_{\nu}$ are 
short hand notation for $ \vec{\alpha}^{(0)}_{\lambda}:=\prod_{p=1}^r \alpha^{(0)}_{\lambda_p}$ for a partition
 $\lambda=[\lambda_1,\cdots,\lambda_l]$ with $\lambda_1\geq\cdots\geq\lambda_l>0$.  The coefficient $\mathcal{O}_{\lambda\nu}$ gives the endmorphism on 
$\otimes_{i=1}^N \cF_i$.  Formally the operator algebra written by  $\mathcal{O}_{\lambda\nu}$
defines the Yangian.  It is directly related to ${\cal W}_N$ algebra with extra $U(1)$ current
algebra as we see in the following.

From the monodromy matrix, one can define mutually commuting operators with 
the transfer matrix $T(u)$, which gives rise to an integrable system.  In the standard construction, it is given as
the trace of $\bT(u)$ over the auxiliary space.
Here such construction does not work, the trace over the Hilbert space $\cF$
diverges in general.  Instead, one may define the transfer matrix
by using the vacuum expectation value of $\bT(u)$:
\ba
T(u):=\ _0\langle\eta|\bT_0(u)|\eta\rangle_0\,.\label{T-def}
\ea
Here we take $u=(\eta_a-\eta)/\sqrt{2}$ with $a=1,\cdots, N$.
Such transfer matrices commute with each other 
from the condition (\ref{R-def2}).
Here is a short computation to prove this:
\ba
T(u)T(u')&=&\ _0\langle \eta|\bT_0(u)|\eta\rangle_0 \ _{0'}\langle \eta'|\bT_{0'}(u\rq{})|\eta'\rangle_{0'}\nn\\
  &=&\ _0\langle \eta|_{0'}\langle \eta'|{\bf R}_{01}(u){\bf R}_{0'1}(u')\dots{\bf R}_{0N}(u){\bf R}_{0'N}(u')
  |\eta\rangle_0|\eta'\rangle_{0'}\nonumber\\
  &=&\ \ _0\langle \eta|_{0'}\langle \eta'|{\bf R}_{00'}(u-u\rq{})^{-1}{\bf R}_{00'}(u-u')
  {\bf R}_{01}(u){\bf R}_{0'1}(u')\dots{\bf R}_{0N}(u){\bf R}_{0'N}(u') |\eta\rangle_0|\eta'\rangle_{0'}\nonumber
\ea
Using the YBE (\ref{YBE}), ${\bf R}_{00'}(u-u'){\bf R}_{0a}(u){\bf R}_{0'a}(u')={\bf R}_{0'a}(u'){\bf R}_{0a}(u){\bf R}_{00'}(u-u')$, 
and the stability conditions for the vacua, (we note that $u-u'=(\eta'-\eta)/\sqrt{2}$)
\ba
\ _0\langle \eta|_{0'}\langle \eta'|{\bf R}_{00'}(u-u')=\ _0\langle \eta|_{0'}\langle \eta'|,\quad 
{\bf R}_{00'}(u-u') |\eta\rangle_0|\eta'\rangle_{0'}= |\eta\rangle_0|\eta'\rangle_{0'},
\ea
the above equation becomes
\ba
T(u)T(u')=\ _0\langle \eta|_{0'}\langle \eta'|{\bf R}_{0'1}(u'){\bf R}_{01}(u)\dots{\bf R}_{0'N}(u'){\bf R}_{0N}(u){\bf R}_{00'}(u-u')
 |\eta\rangle_0|\eta'\rangle_{0'}\nonumber\\
=\ _0\langle \eta|_{0'}\langle \eta'|{\bf R}_{0'1}(u')\cdots{\bf R}_{0'N}(u'){\bf R}_{01}(u)\cdots{\bf R}_{0N}(u)
 |\eta\rangle_0|\eta'\rangle_{0'}=T(u')T(u)\,.\nonumber
\ea

The computation from here is straightforward, coefficients of the $1/u$-expansion of the 
transfer matrix give infinitely many conserved charges.
The existence of such operators implies the integrable structure.

For the one-site case ($N=1$), the first two charges are\footnote{Since there are 
no $\alpha_0^-$ in $r^{(n)}$'s, we will adopt the notation 
$_0\langle0|\dots|0\rangle_0$ for VEV, unless it can be confusing.
We also drop the suffix ($^{(1)}$) in $\alpha_n$ for simplicity.}
\ba
c^{(1)}&:=&\ _0\langle0|r^{(1)}|0\rangle_0=-\rho\sum_{n>0}\alpha_{-n}\alpha_n\,,\\
c^{(2)}&:=&\ _0\langle0|r^{(2)}+\frac{1}{2}(r^{(1)})^2|0\rangle_0=\frac{\rho}{2\sqrt{2}}\sum_{n,m>0}\left(\alpha_{-n}\alpha_{-m}\alpha_{n+m}+\alpha_{-n-m}\alpha_n\alpha_m\right)\nonumber\\
&&+\frac{\rho^2}{2}\left(\sum_{n>0}\alpha_{-n}\alpha_n\right)^2+\frac{\rho^2}{2}\sum_{n>0}n\alpha_{-n}\alpha_n\,.
\ea
Extracting the independent part from the second charge and rescale it, ${\cal H}:=\frac{2\sqrt{2}}{\rho}\left(c^{(2)}-\frac{1}{2}(c^{(1)})^2\right)$, we obtain,
\ba
{\cal H}=\sum_{n,m>0}\left(\alpha_{-n}\alpha_{-m}\alpha_{n+m}+\alpha_{-n-m}\alpha_n\alpha_m\right)-Q\sum_{n>0}n\alpha_{-n}\alpha_n\label{CS}
\ea
with $Q:=-\sqrt{2}\rho$.
This can be identified with the Calogero-Sutherland 
(CS) Hamiltonian written in terms of free boson with the identification
$Q=\sqrt{\beta}-1/\sqrt{\beta}$ where $\beta$ is the coupling constant
of CS \cite{awata1995collective, mimachi1995singular}.
We note that CS is a system of $M$ interacting particles on a circle
with a long range interaction.  The description in terms of free boson
is exact only in the large $M$ limit. 

When $N>1$, these conserved charges are generally the sum of the one-site charge for each site, $c^{(n)}_i$ for the $i$-th site, with additional cross-terms. 
The first two charges are
\ba
c^{(1)} &=&\sum_{i=1}^Nc^{(1)}_i\,,\\
c^{(2)}-\left(c^{(1)}\right)^2&=&\sum_{i=1}^Nc^{(2)}_i+\rho^2\sum_{i<j}\sum_{n>0}n\alpha^{(i)}_{-n}\alpha^{(j)}_n\,.\label{Nc2}
\ea
The higher-rank charges can be computed in the same way. 
\subsection{U(1)$\otimes {\cal W}_N$ Structure in the Integrable System}
We will investigate the algebraic structure of the above integrable system in this subsection. To get prepared, 
we need the commutator results for the one-site case
\ba
\left[\cH, \alpha_n\right]&=& -2n L^{(0)}_n+Q n^2\alpha_n\qquad (n>0)\nonumber\\
\left[\cH, \alpha_{n}\right]&=& -2n L^{(0)}_n-Q n^2\alpha_n\qquad (n<0)\nonumber
\ea
where $L^{(0)}_n$ here is the stress tensor without zero modes for $\alpha_n$. We will denote the above result as $[{\cal H},\alpha_n]=-2nL_n^{(0)}\pm Qn^2\alpha_n$ for short.

In the $N=2$ case, the full Hamiltonian is
\ba\label{fH}
{\cal H}={\cal H}^{(1)}+{\cal H}^{(2)}-2Q\sum_{n>0}n\alpha_{-n}^{(1)}\alpha_n^{(2)}\nonumber
\ea
It coincides with the Hamiltonian for the generalized Calogero-Sutherland
system \cite{Estienne:2011qk, Morozov:2013rma, Shou:2011nu, Shou:2014pxa}.
It is easy to show that
\ba
[{\cal H},\alpha^+_n]=-\frac{2n}{\sqrt{2}}L^{(0)+}_n\pm 2Qn^2\alpha^+_n-\frac{2n}{\sqrt{2}}\left(L_n^{(0)-}+\frac{Q}{\sqrt{2}}n\alpha^-_n\right)\nonumber
\ea
which can be translated to
\ba
{\cal H}=\frac{1}{\sqrt{2}}{\cal H}^+(2\sqrt{2}Q)+\sqrt{2}\sum_n{}'\alpha^+_{-n}\left(L^{(0)-}_n+\frac{Q}{\sqrt{2}}n\alpha^-_n\right)\,.
\ea
The $\phi^-$ part in those formula,
$L_n^-:=L^{(0)-}_n+\frac{Q}{\sqrt{2}}n\alpha^-_n$ is the expression for
Virasoro generator with $c=1-6Q^2$ in the $w$-plane.

For the general $N$,
the commutation relation of the Hamiltonian with $\sum_{n=1}^N \alpha_n^{(a)}$
gives the Virasoro operator which is included in the ${\cal W}_N$ algebra
 (plus the contribution of $U(1)$ factor)
with the central charge,
\ba
c=1+(N-1)(1-Q^2(N^2+N))\,,
\ea
(see for example, chapter 5 of \cite{Kanno:2013aha} for the detail).

We have to mention the intimate relation between the algebra SH${}^c$ \cite{schiffmann2013cherednik}
(a short name for spherical degenerate double affine Hecke algebra)
and the Yangian. The algebra SH${}^c$
consists of an infinite number of generators $D_{r,s}$ with $r,s\in \mathbf{Z}$ with
$r\geq 0$. The algebra contains a parameter $\beta$ and it is reduced to
$\mathcal{W}_{1+\infty}$ algebra when $\beta\rightarrow 1$.
While the algebra is complicated, it can be generated by taking multiple commutators of
three operators $D_{\pm 1,0}\sim \sum_a\alpha^{(a)}_{\mp 1}$, and $D_{0,2}\sim \mathcal{H}$.
It is known that there is a representation of SH${}^c$ by $N$ bosons which
is identical with ${\cal W}_N$ algebra with $U(1)$ factor
(some detail of the correspondence was shown in \cite{schiffmann2013cherednik, Matsuo:2014rba}).

The Yangian for $N$ sites is also described by $N$ bosons.
The defining algebra $\widehat{\mathfrak{gl}(1)}$
is generated by $J_n=\sum_i \alpha^{(i)}_n$ and the zero mode operator
of Virasoro algebra $L_0$.
What we have seen is that it contains the Hamiltonian $\mathcal{H}$
as one of the higher operators.   The definition of the Hamiltonian
with $N$ bosons coincides with $D_{0,2}$ in SH$^c$.
The extra term to define the Hamiltonian (\ref{fH}) comes from the coproduct
in the Yangian.  Similarly, exactly the same term is required to
define the coproduct in  SH${}^c$.
While the construction of general higher order operators
in the Yangian is not obvious, it is natural to expect that
the Yangian and SH$^c$ are the same symmetry.
In this sense, the Yangian construction gives an alternative 
and a systematic definition of ${\cal W}_N$ algebra.
It may be unified with $\mathcal{W}_\infty[\mu]$ symmetry 
\cite{Gaberdiel:2011wb} which
is known to describe the minimal models of ${\cal W}_N$-algebra.
In this sense, it should be also the same symmetry
as SH${}^c$ or the Yangian while the latter contains
extra $U(1)$ factor.

There is another interesting link between ${\cal W}_N$ algebra and MO's R-matrix.
The definition of generators of ${\cal W}_N$ algebra 
(with diagonal $U(1)$ factor) is given in terms
of $N$ bosons through the quantum Miura transformation
\cite{Fateev:1987zh},
\ba
(Q\partial_z -\partial_z \phi^{(1)})\cdots (Q\partial_z -\partial_z \phi^{(N)})
=\sum_{s=0}^N Q^{N-s}W^{(s)}(z)\partial_z^{N-s}\,.
\ea
On the left hand side, there is an ambiguity of ordering
$N$ bosons.  Different ordering gives different but equivalent
set of ${\cal W}$ generators.  MO's R-matrix plays a role of
changing the order,
\ba
{\bf R}_{ab}(Q\partial_z -\partial_z \phi^{(a)}) (Q\partial_z -\partial_z \phi^{(b)})=(Q\partial_z -\partial_z \phi^{(b)}) (Q\partial_z -\partial_z \phi^{(a)}){\bf R}_{ab}\,.\label{miura}
\ea
The proof is straightforward, by using the notation $\phi^{\pm}:=\frac{1}{\sqrt{2}}(\phi^a\pm\phi^b)$, we have
\ba
(Q\partial_z -\partial_z \phi^{(a)}) (Q\partial_z -\partial_z \phi^{(b)})=f(\partial,\phi^+)+\frac{1}{\sqrt{2}}Q\partial^2\phi^--\frac{1}{2}\partial\phi^-\partial\phi^-=f(\partial,\phi^+)-T^-(Q/\sqrt{2})\,.\nn
\ea
$f$ is some function independent of $\phi^-$ and $T^-$ is the Liouville stress tensor on the $\phi^-$ subspace, thus we can see that eq. (\ref{miura}) holds. 

\section{Extension to ${\cal N}=1$ Superconformal Field Theory}\label{s:susy}
In this section, we will introduce an R-matrix respecting the superconformal algebra and build the corresponding integrable system. 
The superconformal algebra is given by the following OPEs between the stress tensor and an $h=3/2$ primary field $T_F(z)$,
\ba
T_B(z)T_B(w)&\sim&\frac{3\hat c}{4(z-w)^4}+\frac{2}{(z-w)^2}T_B(w)+\frac{1}{z-w}\partial T_B(w)\,,\label{B-BOPE}\\
T_B(z)T_F(w)&\sim&\frac{3}{2(z-w)^2}T_F(w)+\frac{1}{z-w}\partial T_F(w)\,,\label{B-FOPE}\\
T_F(z)T_F(w)&\sim&\frac{\hat c}{(z-w)^3}+\frac{2}{z-w}T_B(w)\,,\label{F-FOPE}
\ea
where $\hat c$ is the central charge of super Virasoro algebra 
which is related to the bosonic case as $c=\frac{3\hat c}{2}$.
For notational record, the free fermion in 2D can be expanded as
\ba
\psi(z)=\sum_r\psi_rz^{-r-1/2},\quad
\{\psi_r,\psi_s\}=\delta_{r+s,0}\,.
\ea
For NS sector $r\in \mathbf{Z}+\frac12$ and for Ramond $r\in \mathbf{Z}$.
In the following we focus on the NS sector.
We write the Fock space generated by oscillators  $\alpha_n$, $\psi_r$
from the vacuum $|\eta\rangle$ ($\alpha_n|\eta\rangle =\psi_r |\eta\rangle=0$,
($n,r>0$) and $\alpha_0|\eta\rangle =\eta |\eta\rangle$) as $\cFs(\eta)$.

\subsection{Definition of R-matrix}

The super Virasoro operators are written by free boson and fermion 
as,
\ba
T_B(z)&=&\frac{1}{2}\partial\phi\partial\phi(z)-\rho\partial^2\phi(z)-\frac{1}{2}\psi\partial\psi(z)\nonumber\\
T_F(z)&=&\psi\partial\phi(z)-2\rho\partial\psi(z)
\ea
The R-matrix is defined as an endomorphism on $\cFs\otimes \cFs$.
To describe the Fock space we introduce two sets of free bosons and fermions
$\phi^{(a)}, \psi^{(a)}$ and introduce $\phi^\pm:=\frac{1}{\sqrt2}(\phi^{(1)}\pm\phi^{(2)})$,
$\psi^\pm:=\frac{1}{\sqrt2}(\psi^{(1)}\pm\psi^{(2)})$ as before.
We write the super Virasoro generator in terms of $\phi^-,\psi^-$.
Their mode expansion contains parameters $\rho$ and zero mode $(u)$ of $\phi^-$ as
(again we use the expansion in $w$ plane):
\ba
L_n^{B}(u,\rho)&=&\frac{1}{2}\sum_m{}'\alpha^-_{n-m}\alpha^-_m+\rho n\alpha_n^--u\alpha_n^-+\frac{1}{2}\sum_r(r+1/2):\psi^-_{n-r}\psi^-_r:\nonumber\\
&=:&L_n^{B(0)}+\rho n\alpha_n^--u\alpha_n^-\,,\label{TB-SUSY}\\
L_r^{F}(u,\rho)&=&\sum_{m\neq0}\psi^-_{r-m}\alpha^-_m-u\psi_r^-+2\rho r \psi^-_r\nonumber\\
&=:&L_r^{F(0)}+2\rho r\psi_r^--u\psi_r^-\,,\label{TF-SUSY}
\ea
which gives $\hat c=1-8\rho^2$ SCFT.
We define the reflection operators $R_{\epsilon_1,\epsilon_2}$
by requiring  (\ref{R-def}) for both $L_n^{(B)}, L_r^{(F)}$;
\ba
R_{\epsilon_1,\epsilon_2}(u,\rho)L^B_{n}(u, \rho) =
L^B_{n}(\epsilon_1 u, \epsilon_2 \rho) R_{\epsilon_1,\epsilon_2}(u,\rho)\,,\quad
R_{\epsilon_1,\epsilon_2}(u,\rho)L^F_{n}(u, \rho) =
L^F_{n}(\epsilon_1 u, \epsilon_2 \rho) R_{\epsilon_1,\epsilon_2}(u,\rho)\,.
\ea
We keep also the normalization condition (\ref{R-def2}).
The $\bf{R}$ matrix is defined by $R_{+-}(u,\rho)$ as the bosonic case:
\ba
\mathbf{R}(u)=R_{+-}(u,\rho)=:\sum_{n=0}^\infty R^{(n)} u^{-n}
=:\exp\left(\sum_{n>0} r^{(n)} u^{-n}\right)\,.
\ea
The YBE can be proved as the bosonic case.
%
We can write down the recursion equation for the coefficients $R^{(n)}$ as
\ba
\left[R^{(m)},\alpha^-_n\right]&=&\left[R^{(m-1)},L_n^{B(0)}\right]+\rho n\left\{R^{(m-1)},\alpha^-_n\right\}\nonumber\\
\left[R^{(m)},\psi^-_n\right]&=&\left[R^{(m-1)},L_n^{F(0)}\right]+2n\rho\left\{R^{(m-1)},\psi^-_n\right\}\label{F-recursion}
\ea
The Jacobi identity for these recursion relation gives nontrivial consistency
conditions.  We give an explicit proof for such relations also 
in Appendix \ref{s:Jacobi}. 
It implies that the $R^{(m)}$ can be determined uniquely to the full order.

\subsection{Explicit Form of R-matrix and Conserved charges}
\paragraph{R-matrix}
Let us explore the first few terms of the expansion of R-matrix.
The first set of equations for $r^{(1)}$ is 
\ba
\left[r^{(1)},\alpha_n^-\right]=2\rho n\alpha_n^-\nonumber\\
\left[r^{(1)},\psi_r^-\right]=4\rho r\psi_r^-\,.
\ea
The solution is 
\ba
r^{(1)}=-2\rho\sum_{n>0}\alpha^-_{-n}\alpha^-_n-4\rho\sum_{r>0}r\psi^-_{-r}\psi^-_r\,.\label{r-1-SUSY}
\ea
Substituting this result into equations for $r^{(2)}$, we get
\ba
\left[r^{(2)},\psi^-_r\right]=\left[r^{(1)},\sum_{m\neq0}\psi^-_{r-m}\alpha^-_m\right]=\sum_{m\neq0}(4r-2m)\rho\psi^-_{r-m}\alpha^-_m\,,\label{F-eq-r-2}\\
\left[r^{(2)},\alpha^-_n\right]=\left[r^{(1)},\frac{1}{2}\sum_m{}'\alpha^-_{n-m}\alpha^-_m+\frac{1}{2}\sum_r(r+1/2):\psi^-_{n-r}\psi^-_r:\right]\nonumber\\
=n\rho\sum_m{}'\alpha^-_{n-m}\alpha^-_m+n\rho\sum_r(2r+1):\psi^-_{n-r}\psi^-_r:\,.\label{B-eq-r-2}
\ea
With an identity
\ba
\left[\sum_{m\neq0,r}(2r+1):\psi_{-m-r}\psi_r:\alpha_m,\psi_n\right]=-\sum_{m\neq0}(4n-2m)\psi_{n-m}\alpha_m\,,\nonumber
\ea
we have
\ba
r^{(2)}=-\rho\sum_{n,m>0}\left(\alpha^-_{-n}\alpha^-_{-m}\alpha^-_{n+m}+\alpha^-_{-n-m}\alpha^-_n\alpha^-_m\right)
-\rho\sum_r \sum_{n\neq0}(2r+1):\psi^-_{-n-r}\psi^-_r:\alpha^-_n\,.\label{r-2-SUSY}
\ea

We define the Yangian through the product of the monodromy matrix
(\ref{Tmat}).  The operator algebra is associated with the algebra of
free boson and fermion since the leading order term is,
\ba
{\bf R}_{0i}= 1-\frac{\rho}{u}\left(
\sum_{n>0} (\alpha_{-n}^0-\alpha_{-n}^i) (\alpha_{n}^0-\alpha_n^i)+
2\sum_{r>0} r(\psi_{-r}^0-\psi_{-r}^i) (\psi_{r}^0-\psi_r^i)\right)+\cdots
\ea
This time the underlying algebra is spanned by $K=\sum_{n>0}\alpha_{-n}\alpha_n+\sum_{r>0}2r\psi_{-r}\psi_r$, $\alpha_n$, $\psi_r$ and $1$. 
The non-vanishing commutators in the algebra are
\ba
[K,\alpha_n]=-n\alpha_n\,,\quad[K,\psi_r]=-2r\psi_r\,,\quad[\alpha_n,\alpha_m]=n\delta_{n+m,0}\,,\quad\{\psi_r,\psi_s\}=\delta_{r+s,0}\,.
\ea
Since the algebra contains a minimal supersymmetric system with free boson
and fermion, it may be appropriate to call the new symmetry 
``Yangian associated with superconformal algebra" as the title of this paper \footnote{There are some references on Yangians for fermionic algebra, see for example \cite{superYangian}, in which 
the super Yangian of $gl(M|N)$ was constructed}. 
Again, we can check that the classical Yang-Baxter equation holds for the classical R-matrix (see Appendix \ref{s:cYBE})
\ba
\mathfrak{r}_{ij}=\frac{1}{u_i-u_j}\left(\sum_{n>0}(\alpha_{-n}^i\alpha_n^j+\alpha_n^i\alpha_{-n}^j)+\sum_{r>0}2r(\psi^i_{-r}\psi^j_r-\psi^i_r\psi^j_{-r})-K^{(i)}-K^{(j)}\right)\label{s-classical-r}\,,
\ea
where $K^{(i)}$ stands for the $K$ operator placed on the $i$-th site. 

\paragraph{Conserved charges}
Then we can go further to calculate conserved charges with the same definition of transfer matrix (\ref{T-def}). 
The first charge, is just a sum of $N$ copies of that of a one-site 
system.
\ba
c^{(1)}=-\rho\sum_{a=1}^N\left(
\sum_{n>0}\alpha^{(a)}_{-n}\alpha^{(a)}_n+2\sum_{r>0}r\psi^{(a)}_{-r}\psi^{(a)}_r
\right)\label{c-1-SUSY}
\ea
It may be better to write the second charge in the form of Hamiltonian,
\ba
\cH:=\frac{2\sqrt{2}}{\rho}\left(c^{(2)}-\frac{1}{2}(c^{(1)})^2\right)\,.
\ea
It is divided as a sum over the diagonal and off-diagonal part as,
\ba
\cH=\sum_{a=1}^N\cH^{(a)}+\sum_{a<b} \cV^{(ab)}\,,
\ea
where
\ba
{\cal H}^{(a)}&=&\sum_{n,m>0}\left(\alpha^{(a)}_{-n}\alpha^{(a)}_{-m}\alpha^{(a)}_{n+m}+\alpha^{(a)}_{-n-m}\alpha^{(a)}_n\alpha^{(a)}_m\right)-Q\sum_{n>0}n\alpha^{(a)}_{-n}\alpha^{(a)}_n\nonumber\\
&&+\sum_r\sum_{n \neq 0}(2r+1):\psi^{(a)}_{-r-n}\psi^{(a)}_r:\alpha^{(a)}_n-4Q\sum_{r>0}r^2\psi^{(a)}_{-r}\psi^{(a)}_r\,,\label{supCS}\\
\cV^{(ab)}&=&
-2Q\sum_{n>0}n\alpha^{(a)}_{-n}\alpha^{(b)}_n-8Q\sum_{r>0}r^2\psi^{(a)}_{-r}\psi^{(b)}_r\,.
\ea
For the $N=1$ case, the diagonal part $\cH^{(a)}$ gives a supersymmetric generalization 
of the Calogero-Sutherland Hamiltonian (\ref{CS}).
By construction it is one of the infinite commuting charges defined by 
$c^{(n)}$ ($n=1,2,3,\cdots$).  In this sense, it gives a definition of integrable model
-- a supersymmetric generalization of Calogero-Sutherland model.

\subsection{Comparison with Super CS Model}
A supersymmetric extension of Calogero-Sutherland model
has been developed for some years since \cite{Desrosiers:2001ri}.
They introduced both bosonic and fermionic coordinates $x_i$, $\theta_i$ ($i=1,\cdots, N$)
and defined an integrable system with similar long range interaction.
Let us quote the definition in their recent article \cite{superCS} where
the relation with the superconformal field theory is well-explained.
The operator which plays the role of the Hamiltonian is given as,
($\alpha:=1/\beta$),
\ba
D=\frac{1}{2}\sum_{i=1}^N\alpha x_i^2\partial_{x_i}^2+\sum_{1\leq i\neq j\leq N}\frac{x_ix_j}{x_i-x_j}\left(\partial_{x_i}-\frac{\theta_i-\theta_j}{x_i-x_j}\partial_{\theta_i}\right)\,.
\label{superHam}
\ea
They argued that the eigenstates can not be completely specified
by the Hamiltonian since there are some degeneracy.  To resolve them,
we need another operator $\Delta$,
\ba
\Delta=\sum_{i=1}^N\alpha x_i\theta_i\partial_{x_i}\partial_{\theta_i}+
\sum_{1\leq i\neq j\leq N}\frac{x_i\theta_j+x_j\theta_i}{x_i-x_j}\partial_{\theta_i}\,.
\ea
These operators can be rewritten in the language of free boson and fermion oscillators in the $N\rightarrow\infty$ limit
as in the bosonic case \cite{awata1995collective, mimachi1995singular}.
We introduce the analog of power sum for $n>0$,
\ba
p_n:=\sum_ix^n_i,\ \ \ \ \tilde{p}_n:=\sum_i\theta_ix_i^n.
\ea
We introduce 
bosonic and fermionic oscillators by the correspondences (again $n>0,r>0$)
\ba
a_{-n}\lra\frac{(-1)^{n-1}}{\sqrt{\alpha}}p_n\,,\ \ \ \ \alpha_n\lra n(-1)^{n-1}\sqrt{\alpha}\frac{\partial}{\partial p_n}\,,\nn\\
\psi_{-k}\lra \frac{(-1)^{k-1/2}}{\sqrt{\alpha}}\tilde{p}_{k-1/2}\,,\ \ \ \
 \psi_k\lra(-1)^{k-1/2}\sqrt{\alpha}\frac{\partial}{\partial \tilde{p}_{k-1/2}}\,,\nn
\ea
where the index $r$ for fermion is assumed to be half-integer. 
Based on such correspondence, they derived the correspondence
between super Jack polynomial and the null states of ${\cal N}=1$ superconformal field theory.\cite{superCS}

For the comparison with our model, we derive an oscillator expression for the two
conserved charges \cite{Lapointe:2015kta},
\ba
D\propto \sum_{n,m>0}\left(\alpha_{-n}\alpha_{-m}\alpha_{n+m}+\alpha_{-n-m}\alpha_n\alpha_m\right)+\sum_{n,m>0}2m\left(\psi_{-n-m-1/2}\psi_{m+1/2}\alpha_n
+\psi_{-m-1/2}\psi_{m+n+1/2}\alpha_{-n}\right)\nn\\
-\left(\sqrt{\alpha}-\frac{1}{\sqrt{\alpha}}\right)\left(\sum_{n>0}(n-1)\alpha_{-n}\alpha_n+\sum_{n>0}n(n-1)\psi_{-n-1/2}\psi_{n+1/2}\right)\label{DLM-1}\\
\Delta\propto \sum_{n>0,m\geq 0}\alpha_{-n}\psi_{-m-1/2}\psi_{n+m+1/2}+\sum_{n>0,m\geq0}\psi_{-n-m-1/2}\psi_{m+1/2}\alpha_n
-\left(\sqrt{\alpha}-\frac{1}{\sqrt{\alpha}}\right)\sum_{n>0}n\psi_{-n-1/2}\psi_{n+1/2}\label{DLM-2}
\ea
The Hamiltonian looks similar to the expression (\ref{supCS}) if we identify $Q\equiv\sqrt{\alpha}-\frac{1}{\sqrt{\alpha}}$
but there are some discrepancy.  For example, $\psi\psi\alpha$ term does not have
terms like $\alpha_{-r-s}\psi_r \psi_s $ ($r,s>0$).
It implies that the construction of the supersymmetric generalarization of Calogero-Sutherland
model may not be unique.  Indeed, their construction and ours contain the same
bosonic part and have an infinite number of commuting charges.
In that sense, both constructions define a solvable model with the same
bosonic part.

\section{Conclusion}
In this paper, the Yangian for the ${\cal N}=1$ superconformal algebra
was constructed along the line of proposal \cite{Maulik:2012wi}.
We proved the consistency conditions for the recursion formula
 and derived higher conserved charges,
 the existence of which
implies the integrability of the system
defined in terms of $N$ pairs of free boson and fermion.

Obviously there are many open questions in the model.
In the bosonic case, as we briefly explained, the Yangian is
equivalent to $\mathcal{W}$ algebra and SH${}^c$.
We expected that a similar correspondence exists in our model.
So far, however, in our preliminary study, the super
${\cal W}$ algebra \cite{Inami:1988xy} is not included in the Yangian nor the connection
with the super KP hierarchy \cite{Mathieu:1987xz} could be found.
In this sense, some modification will be necessary to find the link
with the recent developments of super $\mathcal{W}_\infty$ symmetry
(see for example, \cite{Beccaria:2013wqa, Ahn:2013ota}).

\section*{Acknowledgements}
We would like to thank I. Kostov, D. Serban and V. Pasquier for the discussion
at various stages of this work.
RZ would like to thank HIROSE international scholarship for generous 
financial support, Z.Y. Wang for techanical consultation on LaTeX in the first draft and R.Y. Qiu for helpful discussions at the very beginning stage of this project.
YM  would like to thank the hospitality of collegues at IPhT Saclay for their
hospitality during his visit 
and Fondation Math\'ematique Jacques Hadamard for their financial
support. He is also partially supported by Grants-in-Aid for Scientific Research (Kakenhi \#25400246).
\appendix

\section{A Short Review of Yangian}
\label{a:Yangian}
In order to make this paper self-contained, we give a brief summary of the basics of Yangian.
We pick up some relevant parts of  \cite{bernard1992introduction}.

We consider a quantum $su(p)$ Heisenberg spin chain
on a lattice with $N$ points. On each site, we have $su(p)$ algebra with
generators $S^{ab}_j$ ($a,b=1,\cdots, p$, $j=1,\cdots, N$)
which satisfies, $su(p)$ algebra,
\ba
\left[ S^{ab}_j, S^{cd}_k\right]=\delta_{jk}\left(
\delta^{cd}S^{ad}_j-\delta^{ad} S^{cd}_j
\right)\,.
\ea
The Hamiltonian is written in terms of them as,
\ba
H=\sum_{k=1}^N\sum_{ab}S^{ab}_k S^{ba}_{k+1}
\ea
We use the periodic boundary condition and $S^{ab}_{N+1}$
is identified with $S^{ab}_{1}$.

The monodromy matrix is defined by a ordered product of
spin generators,
\ba\label{Tab}
T_{ab}(u) =\sum_{a_1,\cdots, a_{N-1}} \left(1+\frac{h}{u} S_1\right)^{a a_1}\left(1+\frac{h}{u} S_2\right)^{a_1 a_2}
\cdots \left(1+\frac{h}{u} S_N\right)^{a_{N-1} b}\,,
\ea
which satisfies the relation characterizing the monodromy matrix,
\ba
R(u_1-u_2) (T(u_1)\otimes 1)(1\otimes T(u_2))=(1\otimes T(u_2))(T(u_1)\otimes 1)R(u_1-u_2)\,,
\ea
where $R$ matrix is defined as an endmorphism on $\mathbf{C}^p\otimes\mathbf{C}^p$
\ba
R(u)=u-h P,\quad P(x\otimes y):=y\otimes x, \quad (x,y\in \mathbf{C}^p)\,,
\ea
which satisfies the Yang-Baxter relation.

The monodromy matrix can be expanded as,
\ba
T_{ab}(u)=\delta_{ab}+\frac{h}{u}\sum_k S_k^{ab} +\frac{h^2}{u^2} \sum_{j<k}\sum_d S_j^{ad}S_k^{db}+\cdots. 
\ea
The coefficients of the expansion defines a group of generators,
\ba
Q_{ab}^0&=& \sum_k S^{ab}_k\\
Q^1_{ab} &=& \frac{h}{2}\sum_{j<k}\sum_d\left(
S^{ad}_j S^{db}_k-S^{ad}_k S^{db}_j
\right)\,.
\ea
The operators $Q^0_{ab}$ generate $su(p)$ symmetry of the Hamiltonian.
The extra generators $Q^1_{ab}$ give also a(n approximate) symmetry.
They satisfy,
\ba
\left[Q^0_{ab}, Q^0_{cd}\right]&=&\delta_{cd}Q^0_{ad}-\delta_{ad} Q^0_{cd}\\
\left[Q^0_{ab}, Q^1_{cd}\right]&=&\delta_{cd}Q^1_{ad}-\delta_{ad} Q^1_{cd}
\ea
and a complicated nonlinear relations for $\left[Q^1_{ab}, Q^1_{cd}\right]$.
Such algebra generated by the coefficients of the monodromy matrix is
called $su(p)$ Yangian.

A nice property of the Yangian is that it has a well-defined coproduct relation such as,
\ba
\Delta Q^0_{ab}&=& Q^0_{ab}\otimes 1+ 1\otimes Q^0_{ab}\\
\Delta Q^1_{ab}&=& Q^1_{ab}\otimes 1+ 1\otimes Q^1_{ab} +\frac{h}{2}\sum_d
\left(
Q^0_{ad}\otimes Q^0_{db}-Q^0_{db}\otimes Q^0_{ad}
\right)\,,
\ea
which comes from the product of two monodromy matrices.

\section{Checking the classical Yang-Baxter equation}\label{s:cYBE}
As we are free to rescale the classical R-matrix with an overall factor, let us multiply a factor of $\sqrt{(u_1-u_2)(u_1-u_3)(u_2-u_3)}$ to (\ref{classical-r}) or (\ref{s-classical-r}) for convenience. 

In the bosonic case, we have
\ba
\left[\mathfrak{r}_{12},\mathfrak{r}_{13}\right]=-(u_2-u_3)\sum_{n>0}n\alpha_{-n}\otimes1\otimes\alpha_n+(u_2-u_3)\sum_{n>0}n\alpha_n\otimes1\otimes\alpha_{-n}\nn\\
+(u_2-u_3)\sum_{n>0}n\alpha_{-n}\otimes\alpha_n\otimes1-(u_2-u_3)\sum_{n>0}n\alpha_n\otimes\alpha_{-n}\otimes1\nn\\
-(u_2-u_3)\sum_{n>0}n1\otimes\alpha_n\otimes\alpha_{-n}+(u_2-u_3)\sum_{n>0}n1\otimes\alpha_{-n}\otimes\alpha_n\,.
\ea
The other two terms are obtained by permutation.  The sum of three terms vanishes identically.

The fermionic part in the superconformal case cancels in the similar way, 
but this time we have to pay attention to the anticommutative nature of fermionic operators. 
For instance,
\ba
\left[\sum_{r>0}\psi^1_{-r}\psi^2_r,\sum_{s>0}\psi^1_s\psi^3_{-s}\right]=\sum_{r,s>0}\left(\psi^1_{-r}\psi^2_r\psi^1_s\psi^3_{-s}-\psi^1_s\psi^3_{-s}\psi^1_{-r}\psi^2_r\right)\nn\\
=\sum_{r,s>0}\left(-\psi^1_{-r}\psi^1_s\psi^2_r\psi^3_{-s}-\psi^1_s\psi^1_{-r}\psi^2_r\psi^3_{-s}\right)=-\sum_{r>0}\psi^2_r\psi^3_{-r}\,.\nn
\ea
Therefore, the fermionic part of the commutator in $\left[\mathfrak{r}_{12},\mathfrak{r}_{13}\right]$
is given by
\ba
-(u_2-u_3)\sum_{r>0}4r^2\psi_{-r}^1\psi_r^3-(u_2-u_3)\sum_{r>0}4r^2\psi_r^1\psi_{-r}^3\nn\\
+(u_2-u_3)\sum_{r>0}4r^2\psi_{-r}^1\psi_r^2+(u_2-u_3)\sum_{r>0}4r^2\psi_r^1\psi_{-r}^2\nn\\
+(u_2-u_3)\sum_{r>0}4r^2\psi_r^2\psi_{-r}^3+(u_2-u_3)\sum_{r>0}4r^2\psi_{-r}^2\psi_r^3\,.
\ea
With other two commutators obtained by permutation, we can confirm the cancellation 
fermionic part again.

\section{Proof of Jacobi identity}\label{s:Jacobi}
The set of recursion equations (\ref{recursion}) is in general not self-consistent, it is only when the following Jacobi identity hlods for any $i$, $n$ and $m$ that the existence of the R-matrix is ensured. 
\ba
[[R^{(i)},\alpha^-_n],\alpha^-_m]=[[R^{(i)},\alpha^-_m],\alpha^-_n].\nonumber
\ea
Let us give the proof here. The calculation is complicated but the strategy is simple: using the recursion equation to rewrite the expression for both sides of (\ref{Jacobi}) with lower level of $R^{(i)}$. 
\ba
(l.h.s.)=[[R^{(i-1)},L_n^{(0)}],\alpha_m^-]+\rho n[R^{(i-1)}\alpha^-_n+\alpha^-_nR^{(i-1)},\alpha^-_m]\nonumber\\
=-[[L_n^{(0)},\alpha_m^-],R^{(i-1)}]-[[\alpha^-_m,R^{(i-1)}],L_n^{(0)}]+2\rho n^2R^{(i-1)}\delta_{n+m,0}\nonumber\\
+\rho n[R^{(i-1)},\alpha_m^-]\alpha_n^-+\rho n\alpha_n^-[R^{(i-1)},\alpha_m^-]\nonumber\\
=-m[R^{(i-2)},L^{(0)}_{n+m}]+[[R^{(i-2)},L_m^{(0)}],L_n^{(0)}]+\rho m^2R^{(i-2)}\alpha^-_{n+m}+\rho m^2\alpha^-_{n+m}R^{(i-2)}\nonumber\\
+\rho m[R^{(i-2)},L_n^{(0)}]\alpha_m^-+\rho m\alpha_m^-[R^{(i-2)},L_n^{(0)}]
+\rho n[R^{(i-2)},L_m^{(0)}]\alpha^-_n+\rho n\alpha^-_n[R^{(i-2)},L_m^{(0)}]\nonumber\\
-\rho(n+m)mR^{(i-2)}\alpha^-_{n+m}-\rho(n+m)m\alpha^-_{n+m}R^{(i-2)}\nonumber\\
+\rho^2 nm\left(R^{(i-2)}\alpha^-_m\alpha^-_n+\alpha^-_mR^{(i-2)}\alpha^-_n+\alpha^-_n(R^{(i-2)}\alpha^-_m+\alpha^-_n\alpha^-_mR^{(i-2)}\right)\nonumber\\
+2\rho n^2R^{(i-1)}\delta_{n+m,0}\nonumber
\ea
This deformation only holds for $n\neq -m$, and in that case, the second and the fourth line of the final result is symmetric about $n$ and $m$, and the last line is meaningless. 
The last two terms in the first line and the third line combine into $-\rho nmR^{(i-2)}\alpha^-_{n+m}-\rho nm\alpha^-_{n+m}R^{(i-2)}$, and is also symmetric about $n$ and $m$. 
The remaining two terms are actually also symmetric about $n$ and $m$ due to the following fact
\ba
-m[R^{(i-2)},L^{(0)}_{n+m}]+[[R^{(i-2)},L_m^{(0)}],L_n^{(0)}]=-m[R^{(i-2)},L^{(0)}_{n+m}]+[[R^{(i-2)},L_n^{(0)}],L_m^{(0)}]+[R^{(i-2)},[L_m^{(0)},L_n^{(0)}]]\nonumber\\
=[[R^{(i-2)},L_n^{(0)}],L_m^{(0)}]+(m-n)[R^{(i-2)},L^{(0)}_{m+n}]-m[R^{(i-2)},L^{(0)}_{n+m}]=[[R^{(i-2)},L_n^{(0)}],L_m^{(0)}]-n[R^{(i-2)},L^{(0)}_{n+m}]\nonumber
\ea
When $n+m=0$, $[L^{(0)}_n,\alpha^-_m]=0$. Therefore
\ba
(l.h.s)=-[[\alpha^-_m,R^{(i-1)}],L_n^{(0)}]+2\rho n^2R^{(i-1)}\delta_{n+n,0}+\rho n[R^{(i-1)},\alpha_m^-]\alpha_n^-+\rho n\alpha_n^-[R^{(i-1)},\alpha_m^-]\nonumber\\
=[[R^{(i-2)},L_m^{(0)}],L_n^{(0)}]+\rho m[R^{(i-2)}\alpha_m^-+\alpha_m^-R^{(i-2)},L_n^{(0)}]+2\rho n^2R^{(i-1)}\delta_{n+n,0}\nonumber\\
+\rho n[R^{(i-2)},L_m^{(0)}]\alpha^-_n+\rho n\alpha^-_n[R^{(i-2)},L_m^{(0)}]\nonumber\\
+\rho^2nm\left(R^{(i-2)}\alpha^-_m\alpha^-_n+\alpha^-_mR^{(i-2)}\alpha^-_n+\alpha^-_n(R^{(i-2)}\alpha^-_m+\alpha^-_n\alpha^-_mR^{(i-2)}\right)\nonumber\\
=[[R^{(i-2)},L_m^{(0)}],L_n^{(0)}]+2\rho n^2R^{(i-1)}\delta_{n+n,0}\nonumber\\
+\rho m[R^{(i-2)},L_n^{(0)}]\alpha_m^-+\rho m\alpha_m^-[R^{(i-2)},L_n^{(0)}]
+\rho n[R^{(i-2)},L_m^{(0)}]\alpha^-_n+\rho n\alpha^-_n[R^{(i-2)},L_m^{(0)}]\nonumber\\
+\rho^2nm\left(\frac{1}{2}R^{(i-2)}\{\alpha^-_m,\alpha^-_n\}+\alpha^-_mR^{(i-2)}\alpha^-_n+\alpha^-_n(R^{(i-2)}\alpha^-_m+\{\alpha^-_n,\alpha^-_m\}R^{(i-2)}\right)\nonumber
\ea
The only term which is not manifestly symmetric about $n$ and $m$ is the first one, $[[R^{(i-2)},L_m^{(0)}],L_n^{(0)}]$. Actually
\ba
[[R^{(i-2)},L_m^{(0)}],L_n^{(0)}]=[[R^{(i-2)},L_n^{(0)}],L_m^{(0)}]+[R^{(i-2)},[L_m^{(0)},L_n^{(0)}]]\nonumber\\
=[[R^{(i-2)},L_n^{(0)}],L_m^{(0)}]+2m[R^{(i-2)},L_0^{(0)}]\nonumber
\ea
However, the last term of the last line actually vanishes. This can be seen from (\ref{recursion}), for $n=0$, we have
\ba
[R^{(i-1)},L^{(0)}_0]=[R^{(i)},\alpha^-_0]\nonumber
\ea
$R^{(i)}$ must commute with $\alpha^-_0$ or it will contain $\rho_0^-$, 
which means that ${\bf R}(u)$ 
will shift the vacuum. 
Thus any $R^{(i)}$ must commute with $L_0^{(0)}$.

Combining discussions of the two cases above, we see that $[[R^{(i)},\alpha^-_n],\alpha^-_m]$ is symmetric on $n\leftrightarrow m$, thus (\ref{Jacobi}) holds
\footnote{For $i=1$, there is no $R^{(-1)}$, however, we can consider $R^{(-1)}=0$ here, because it satisfies the recursion equation (\ref{recursion}) for $R^{(0)}=1$.}. 
This is a useful proof for the existence of the R-matrix to full order, while ${\bf R}(u)$ can be solved explicitly to only the very first few orders. 

Let us move to the superconformal case. We will show the Jacobi identities $[[R^{(i)},\alpha_n],\psi_m]=[[R^{(i)},\psi_m],\alpha_n]$ and $\{[R^{(i)},\psi_n],\psi_m\}=-\{[R^{(i)},\psi_m],\psi_n\}$ 
in the following. 
The upper indices $^-$ will be omitted here for convenience. The proof of $[[R^{(i)},\alpha_n],\alpha_m]=[[R^{(i)},\alpha_m],\alpha_n]$ is completely the same as has been done in 
the above.

Let us first focus on the former identity. 
\begin{eqnarray}
  [[R^{(i)},\alpha_n],\psi_m]
  =[[R^{(i-2)},L_m^{F(0)}],L_n^{B(0)}]+2\rho m[R^{(i-2)}\psi_m+\psi_mR^{(i-2)},L_n^{B(0)}]\nonumber\\
  +\left(\frac{n}{2}+m\right)[\psi_{n+m},R^{(i-1)}]+2\rho^2 nm\left(R^{(i-2)}\psi_m+\psi_mR^{(i-2)}\right)\alpha_n\nonumber\\
  +\rho n[R^{(i-2)},L_m^{F(0)}]\alpha_n+\rho n\alpha_n[R^{(i-2)},L_m^{F(0)}]\nonumber\\
  +2\rho^2 nm\alpha_n\left(R^{(i-2)}\psi_m+\psi_mR^{(i-2)}\right)\nonumber
\end{eqnarray}
\begin{eqnarray}
  [[R^{(i)},\psi_m],\alpha_m]
  =[[R^{(i-2)},L_n^{B(0)}],L_m^{F(0)}]+\rho n[R^{(i-2)}\alpha_n+\alpha_nR^{(i-2)},L_m^{F(0)}]\nonumber\\
  +n[\psi_{n+m},R^{(i-1)}]+2\rho m[R^{(i-2)},L_n^{B(0)}]\psi_m+2\rho m\psi_m[R^{(i-2)},L_n^{B(0)}]\nonumber\\
  +2\rho^2 nm\left(R^{(i-2)}\alpha_n+\alpha_nR^{(i-2)}\right)\psi_m++2\rho^2 nm\psi_m\left(R^{(i-2)}\alpha_n+\alpha_nR^{(i-2)}\right)\nonumber
\end{eqnarray}
\begin{eqnarray}
  \Rightarrow[[R^{(i)},\alpha_n],\psi_m]-[[R^{(i)},\psi_m],\alpha_n]
  =[(n/2-m)L_{n+m}^{F(0)},R^{(i-2)}]+(m-n/2)[L_{n+m}^{F(0)},R^{(i-2)}]\nonumber\\
  +(n/2-m)2\rho(n+m)(\psi_{n+m}R^{(i-2)}+R^{(i-2)}\psi_{n+m})\nonumber\\
  -\rho n^2R^{(i-2)}\psi_{n+m}-\rho n^2\psi_{n+m}R^{(i-2)}\nonumber\\
  +2\rho mR^{(i-2)}(n/2+m)\psi_{n+m}+2\rho m(n/2+m)\psi_{n+m}R^{(i-2)}\nonumber\\
  =0\nonumber
\end{eqnarray}
The second identity should be discussed in two cases. When $n\neq m$, 
\begin{eqnarray}
  \{[R^{(i)},\psi_n],\psi_m\}
  =-\{[R^{(i-2)},L_m^{F(0)}],L_n^{F(0)}\}+2\rho m[R^{(i-2)},L_n^{F(0)}]\psi_m-2\rho n[R^{(i-2)},L_m^{F(0)}]\psi_n\nonumber\\
  -2\rho m\psi_m[R^{(i-2)},L_n^{F(0)}]+2\rho n\psi_n[R^{(i-2)},L_m^{F(0)}]\nonumber\\
  +4\rho^2nm\psi_n(R^{(i-2)}\psi_m+\psi_mR^{(i-2)})-4\rho^2nm(R^{(i-2)}\psi_m+\psi_mR^{(i-2)})\psi_n\nonumber\\
  -2\rho mR^{(i-2)}\alpha_{n+m}-2\rho m\alpha_{n+m}R^{(i-2)}\nonumber\\
  +[R^{(i-2)},L_{n+m}^{B(0)}]+\rho(n+m)R^{(i-2)}\alpha_{n+m}+\rho(n+m)\alpha_{n+m}R^{(i-2)}\nonumber
\end{eqnarray}
Terms not apparently antisymmetric can be grouped into two parts, 
\begin{eqnarray}
  -\{[R^{(i-2)},L_m^{F(0)}],L_n^{F(0)}\}+[R^{(i-2)},L_{n+m}^{B(0)}]-\{[R^{(i-2)},L_n^{F(0)}],L_m^{F(0)}\}+[R^{(i-2)},L_{n+m}^{B(0)}]\nonumber\\
  =-2[R^{(i-2)},L_{n+m}^{B(0)}]+[R^{(i-2)},L_{n+m}^{B(0)}]+[R^{(i-2)},L_{n+m}^{B(0)}]=0\nonumber
\end{eqnarray}
\begin{eqnarray}
  4\rho^2nm(\psi_nR^{(i-2)}\psi_m+\psi_n\psi_mR^{(i-2)}-R^{(i-2)}\psi_m\psi_n-\psi_mR^{(i-2)}\psi_n)\nonumber\\
  +4\rho^2nm(\psi_mR^{(i-2)}\psi_n+\psi_m\psi_nR^{(i-2)}-R^{(i-2)}\psi_n\psi_m-\psi_nR^{(i-2)}\psi_m)\nonumber\\
  =4\rho^2nm(\{\psi_n,\psi_m\}R^{(i-2)}-R^{(i-2)}\{\psi_n,\psi_m\})=0\nonumber
\end{eqnarray}
If $n=m$, we have to be a little more careful as $\{L_n^{F(0)},\psi_m\}=0$, 
\begin{eqnarray}
  \{[R^{(i)},\psi_n],\psi_m\}
  =-\{[R^{(i-2)},L_m^{F(0)}],L_n^{F(0)}\}+4\rho nR^{(i-1)}\delta_{n+m,0}-4\rho^2nm\psi_mR^{(i-2)}\psi_n+4\rho^2nm\psi_nR^{(i-2)}\psi_m\nonumber\\
  +2\rho m[R^{(i-2)},L_n^{F(0)}]\psi_m-2\rho m\psi_m[R^{(i-2)},L_n^{F(0)}]-2\rho n[R^{(i-2)},L_m^{F(0)}]\psi_n+2\rho n\psi_n[R^{(i-2)},L_m^{F(0)}]\nonumber
\end{eqnarray}
The first term can be transformed to
\begin{eqnarray}
  -\{[R^{(i-2)},L_m^{F(0)}],L_n^{F(0)}\}=\{[R^{(i-2)},L_n^{F(0)}],L_m^{F(0)}\}-[R^{(i-2)},\{L_n^{F(0)},L_m^{F(0)}\}]\nonumber\\
  =\{[R^{(i-2)},L_n^{F(0)}],L_m^{F(0)}\}-2[R^{(i-2)},L_{n+m}^{B(0)}]\nonumber
\end{eqnarray}
As discussed previously, the last term vanishes. Therefore $\{[R^{(i)},\psi_n],\psi_m\}$ is antisymmetric about $n$ and $m$.
\section{Computation of Higher-Rank Coefficients}\label{s:r-3}
In this section, we will show the explicit computation of $r^{(3)}$ in MO's case and try to convince the reader that principally it is possible to do this kind of computations for 
higher-rank coefiicients despite that the difficulty increases exponentially, especially difficult in the ${\cal N}=1$ superconformal case. 
Again the superscript $^-$ will be omitted here. 

The equation for $r^{(3)}$ is
\ba
[r^{(3)},\alpha_n]=[r^{(2)},L_n^{(0)}]+\frac{1}{2}[(r^{(1)})^2,L_n^{(0)}]+\rho n\left(r^{(2)}+\frac{1}{2}(r^{(1)})^2\right)\alpha_n+\rho n\alpha_n\left(r^{(2)}+\frac{1}{2}(r^{(1)})^2\right)
\nonumber\\
-\frac{1}{2}[r^{(1)}r^{(2)},\alpha_n]-\frac{1}{2}[r^{(2)}r^{(1)},\alpha_n]-\frac{1}{3!}[(r^{(1)})^3,\alpha_n]\nonumber\\
=[r^{(2)},L_n^{(0)}]-\frac{1}{6}\rho n[r^{(1)},\alpha_n]r^{(1)}+\frac{1}{6}\rho nr^{(1)}[r^{(1)},\alpha_n]\nonumber
\ea
Using (this also holds for $n=0$)
\ba
[r^{(2)},L_n^{(0)}]=[r^{(2)},\frac{1}{2}\sum_{m}{}'\alpha_{n-m}\alpha_m]=2\rho\sum_{m\neq0,n}(n-m)L_{n-m}^{(0)}\alpha_m+\rho\sum_{m\neq n/2}m(n-m)\alpha_n\nn\\
=2\rho\sum_{m\neq0,n}(n-m)L_{n-m}^{(0)}\alpha_m-\rho\sum_{m\neq n/2}{}'m^2\alpha_n=2\rho\sum_{m\neq0,n}(n-m):L_{n-m}^{(0)}\alpha_m:\nn
\ea
we get
\ba
[r^{(3)},\alpha_n]=2\rho\sum_{m\neq0,n}(n-m):L_{n-m}^{(0)}\alpha_m:+\frac{2}{3}\rho^3n^3\alpha_n\nn
\ea
Note that by changing dummy variables, we can obtain the identity
$3\sum_{m,l}{}'m:\alpha_{n-m-l}\alpha_m\alpha_l:=\sum_{m,l}{}'n:\alpha_{n-m-l}\alpha_m\alpha_l:\nn$, thus 
the equation for $r^{(3)}$ can be finally simplified to
\ba
[r^{(3)},\alpha_n]=\frac{4}{3}\rho n\sum_{m\neq0,n}:L_{n-m}^{(0)}\alpha_m:+\frac{2}{3}\rho^3n^3\alpha_n\nn
\ea
Now it can be solved as
\ba
r^{(3)}=-\frac{1}{3}\rho\sum_{n,m}{}':L_{n-m}^{(0)}\alpha_{m}\alpha_{-n}:-\frac{1}{3}\rho^3\sum_nn^2:\alpha_{-n}\alpha_n:\nn\\
=-\frac{2}{3}\rho\sum_{l,m,n>0}(\alpha_{-l}\alpha_{-m}\alpha_{-n}\alpha_{m+l+n}+\alpha_{-m-n-l}\alpha_l\alpha_m\alpha_n)\nn\\
-\rho\sum_{\substack{m,n,k,l>0\\m+n=k+l}}\alpha_{-k}\alpha_{-l}\alpha_m\alpha_n-\frac{2}{3}\rho^3\sum_{n>0}n^2\alpha_{-n}\alpha_n\label{r-3}
\ea

$r^{(4)}$ is also not hard to compute, we will only show the result here, readers can check the expression by themself.
\ba
r^{(4)}=-\frac{\rho}{10}\sum_{n,m,l,k}{}':\alpha_{-n-m-l-k}\alpha_k\alpha_l\alpha_m\alpha_n:-\frac{1}{2}\rho^3\sum_{n,m}{}':\alpha_{-n-m}\alpha_n\alpha_m:\label{r-4}
\ea

$r^{(3)}$ in the superconformal case can be computed in the same way, the result is
\ba
r^{(3)}=
\frac{1}{6}\rho\sum_{n,m,l}{}':\alpha_{-l-m-n}\alpha_l\alpha_m\alpha_n:-\frac{2}{3}\rho^3\sum_{n>0}n^2\alpha_{-n}\alpha_n\nn\\
-2\rho\sum_{n,m\neq0,r}(n+r):\psi_{-n-m-r}\psi_r\alpha_m\alpha_n:\nn\\
+\rho\sum_{m,s,t}ms:\psi_{t-m}\psi_{m-s}\psi_s\psi_{-t}:-\frac{16}{3}\rho^3\sum_{r>0}r^3\psi_{-r}\psi_r\label{SUSY-r-3}
\ea

\bibliography{bib}
\end{document}